 \documentclass[final,5p,times,twocolumn]{elsarticle}
 \biboptions{numbers,sort&compress}
\usepackage[table]{xcolor}
\usepackage{graphicx,color,amsmath,amssymb}
\usepackage[colorlinks=true,linkcolor=blue,citecolor=blue,urlcolor=blue]{hyperref}
\newcommand{\npart}{N_{\rm part}}
\newcommand{\dd}{\mathrm{d}}
\newcommand{\pT}{p_{\rm T}}
\newcommand{\Tf}{T_{\rm f}}
\newcommand{\Tm}{T_{\rm m}}
\newcommand{\tm}{\tau_{\rm m}}
\newcommand{\ie}{\textit{i.e.}}
\newcommand{\eg}{\textit{e.g.}}
\newcommand{\raa}{R_{\rm AA}}
\newcommand{\bbb}{b\bar{b}}
\newcommand{\Neq}{N_{Y}^{\rm eq}}

\journal{Physics Letters B}

\begin{document}

\begin{frontmatter}



\title{Bottomonium transport in a strongly coupled quark-gluon plasma}


\author[first]{Biaogang Wu}
\author[first]{Ralf Rapp}
\affiliation[first]{organization={
      Cyclotron Institute and Department of Physics and Astronomy,
   Texas A\&M University},
            addressline={}, 
            city={College Station},
            postcode={77843-3366}, 
            state={Texas},
         country={USA}}

\begin{abstract}
Quarkonium production in high-energy heavy-ion collisions remains a key probe of the quark-gluon plasma formed in these reactions,
but the development of a fully integrated nonperturbative approach remains a challenge. Toward this end, we set up a semiclassical transport approach that combines nonperturbative reaction rates rooted in lattice-constrained $T$-matrix interactions with a viscous hydrodynamic medium evolution. Bottomonium suppression is computed along trajectories in the hydrodynamic evolution while regeneration is evaluated via a rate equation extended to a medium with spatial gradients. The much larger reaction rates compared to previous calculations markedly enhance both dissociation and regeneration processes. This, in particular, requires a reliable assessment of bottomonium equilibrium limits and of the non-thermal distributions of the bottom quarks transported through the expanding medium. Within current uncertainties our approach can describe the centrality dependence of bottomonium yields measured in Pb-Pb ($\sqrt{s_{_{\rm NN}}}$=5.02\,TeV) collisions at the LHC, while discrepancies are found at large transverse momenta.   
\end{abstract}



\begin{keyword}
Ultra-relativistic Heavy-Ion Collisions \sep Quark-Gluon Plasma \sep Bottomonia



\end{keyword}

\end{frontmatter}




\section{Introduction}
\label{sec:introduction}
The fundamental force between a heavy quark and its antiquark is well established by now, essentially following the Cornell potential~\cite{Eichten:1974af} that was originally introduced on phenomenological grounds to describe charmonium and bottomonium spectra.
The long-distance linear part of this potential can be considered as one of the most direct manifestations of confinement in the spectroscopy of hadrons. As such, its medium modifications in a quark-gluon plasma (QGP) play a pivotal role in understanding the fate of the confining force in QCD matter, and how this force affects the QGP's properties. 
Heavy-quarkonium production in ultra-relativistic heavy-ion collisions (URHICs) is therefore a prime observable to address these questions.
Modern transport approaches have reached a fair description of pertinent observables from SPS via RHIC to the LHC to date~\cite{Rapp:2008tf,Braun-Munzinger:2009dzl,Liu:2010ej,Mocsy:2013syh, Ferreiro:2012rq,Liu:2015izf, Chen:2017duy,Wolschin:2020kwt,Brambilla:2023hkw,Wu:2024gil,Song:2024got, Daddi-Hammou:2025hdz}.
However, the microscopic ingredients to these calculations, as well as the medium evolution model required to carry out the transport simulations, vary considerably~\cite{Andronic:2024oxz}. In particular, the connection to the strong-coupling nature of the QGP, as has been deduced from transport calculations of individual charm and bottom quarks (and their hadronization)~\cite{Rapp:2018qla,ALICE:2021rxa}, has not yet been established. Furthermore, constraints on the quarkonium transport parameters that can be obtained from first-principles lattice-QCD computations have not been systematically implemented either. 

In the present work we progress in this direction by implementing our recently developed nonperturbative reaction 
rates~\cite{Wu:2025hlf,Tang:2023tkm} into a state-of-the-art viscous hydrodynamic simulation of Pb-Pb collisions at the LHC~\cite{Strickland:2013uga}. Both components represent a
notable advance over our previous work where the quarkonium reaction rates, albeit based on nonperturbative binding energies, were
calculated with a perturbative coupling to the medium and applied in a schematic fireball expansion. The reaction rates are based on nonperturbative $T$-matrix interactions that have been thoroughly constrained by recent lattice-QCD data in both heavy- and light-parton sectors~\cite{Tang:2023tkm}.
Pertinent transport coefficients for heavy-quark (HQ) diffusion have been found to be in the range required by open heavy-flavor (HF) phenomenology at
the LHC~\cite{ALICE:2021rxa} and agree fairly well with the HQ diffusion coefficient from lattice-QCD~\cite{Altenkort:2023eav}.  In this way, the nonperturbative rates and the viscous-hydro evolution used in the present work, constitute a largely parameter-free combination of the two main building blocks of a heavy-quarkonium transport model in URHICs, and both components embody the notion of a strongly coupled QGP.
That being said, there remain significant uncertainties in our current calculations of observables that we will elaborate on. 
Specifically, regeneration processes, which couple to the open HF sector (both its kinetics and chemistry), contribute to these uncertainties in several respects. In our applications to experiment, we therefore focus on bottomonium ($Y$) observables, where regeneration yields are believed to be smaller than for charmonia and thus are expected to provide better sensitivity to the inelastic reaction rates.   


The remainder of this paper is organized as follows: In Sec.~\ref{sec:hydro} we lay out our implementation of dissociation reactions into trajectories obtained from relativistic hydrodynamics including pre-equilibrium effects associated with nuclear shadowing 
and bottomonium formation times. 
In Sec.~\ref{sec:kinetic} we set up the rate equation framework for use in hydrodynamics that enables the computation of regeneration reactions in the presence of a spatially non-uniform medium evolution, including escape effects of bottom quarks and their thermal relaxation.  
In Sec.~\ref{sec:data} we compare the results of our updated framework to available bottomonium data in Pb-Pb collisions at the LHC,  and in Sec.~\ref{sec:sum}  we summarize and conclude.

\section{$Y$ Suppression along trajectories from hydrodynamics}
\label{sec:hydro}

\begin{figure}[t]
	\centering 
        \includegraphics[width=0.405\textwidth]{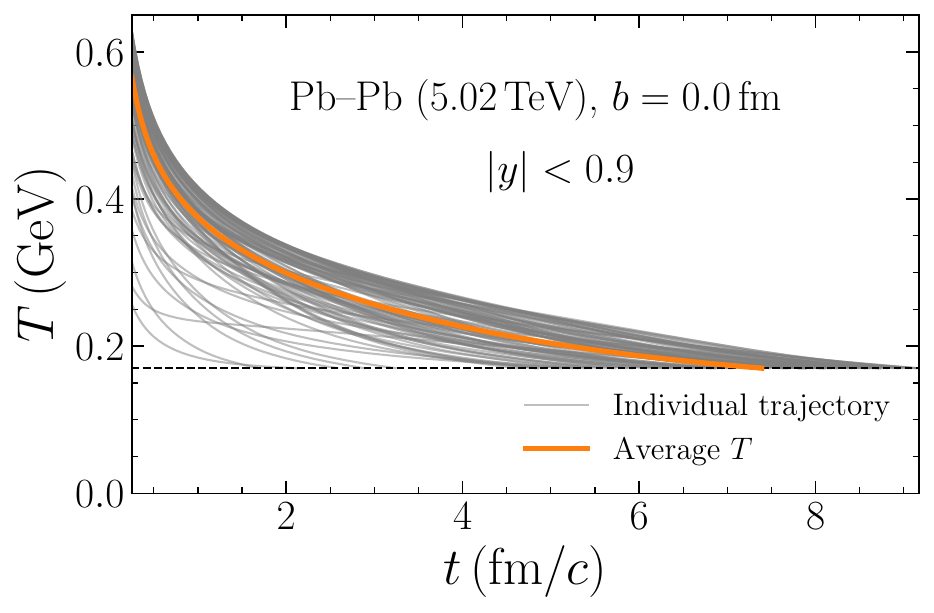}
	\hfill
	\includegraphics[width=0.405\textwidth]{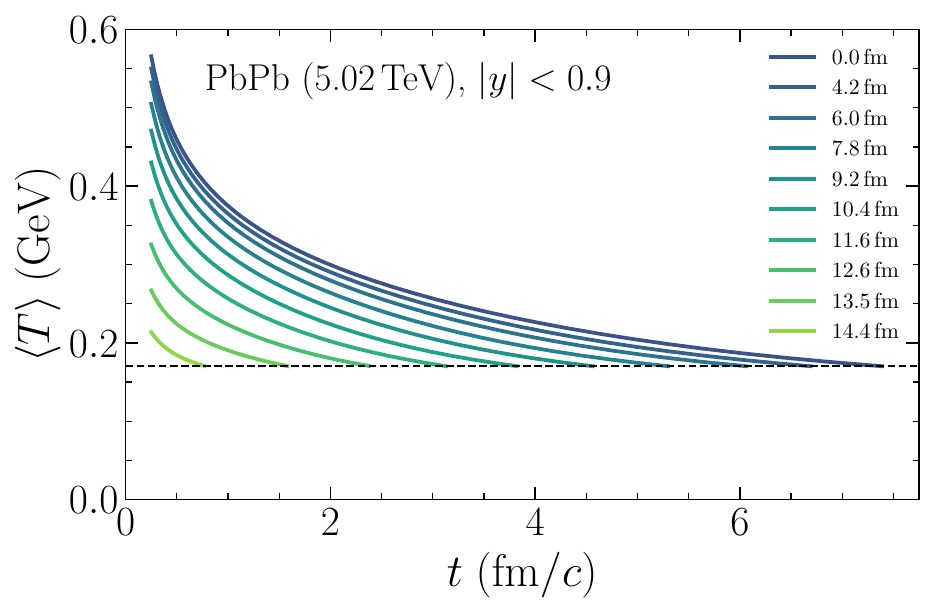}	
	\caption{
	   Time evolution of temperature along $Y$ trajectories in viscous hydrodynamics for Pb-Pb collisions at
           $\sqrt{s_{_{\mathrm{NN}}}} = 5.02\,$TeV in a rapidity window $|y| < 0.9$.
    \textbf{Top panel:} for individual trajectories (gray lines) and their average (thick orange curve) in central collisions ($b = 0$~fm).
    \textbf{Bottom panel:} averages for different centralities (for impact parameters from 0 to 14.4~fm). The horizontal dashed line indicates $\Tf$=0.17\,GeV.
     } 
\label{fig:traj}
\end{figure}

For the bulk medium evolution we employ 3+1-dimensional anisotropic hydrodynamics simulations based on a lattice-QCD equation of state~\cite{Borsanyi:2010cj,HotQCD:2014kol} and parameters (\eg, viscosities) tuned to reproduce experimental data for soft-hadron production~\cite{Strickland:2013uga,Alqahtani:2020paa}.
Focusing on the Pb-Pb (5.02\,TeV) system, we initialize $Y$ trajectories 
in space by an optical Glauber model using a binary-collision profile of hard production, restricted to the rapidity interval of interest.
The initial $Y$ momenta are sampled from an azimuthally symmetric distribution in transverse momentum ($\pT$) that reproduces measured spectra in proton-proton ($pp$) collisions.
Due to their large mass and relatively small binding energies we assume straight-line motion, \ie, neglect elastic rescattering.
Each state, $Y$=$\Upsilon(1S,2S,3S)$, $\chi_b(1P,2P)$, is then dissociated along its trajectory according to its individual momentum- and temperature-dependent rates~\cite{Wu:2025hlf} according to the loss term of the rate equation, 
\begin{equation}
\frac{\dd N_{\Upsilon}(\tau;p)}{\dd\tau} = - \Gamma_{\Upsilon}(T(\tau),p) \ N_{\Upsilon}(\tau;p) \ , 
\end{equation}
which we integrate down to a final temperature of $\Tf$=170\,MeV (lower values do not significantly affect our final results).
The trajectories exhibit a considerable range in temperature and lifetime (cf.~the upper panel of Fig.~\ref{fig:traj} for central Pb-Pb collisions), while the trajectory average is rather similar to previous fireball calculations~\cite{Du:2017qkv}, albeit with a somewhat smaller lifetime in the hydrodynamic environment (cf.~the lower panel of Fig.~\ref{fig:traj} for different collision centralities).

Cold-nuclear-matter effects at the LHC are chiefly expected from nuclear shadowing of the parton distribution functions (PDFs) in the incoming nuclei, which affects the initial hard production of bottomonia. We implement this as in our previous work~\cite{Wu:2024gil}, by adopting a suppression of 20-30\% on the integrated $Y$ production at mid-rapidity and 25-35\% at forward rapidity, also compatible with the studies of Ref.~\cite{Du:2017qkv} where 25\% and 30\% was used, respectively. The $\pT$ dependence is taken from
nuclear PDFs within the EPPS21 fit~\cite{Eskola:2021nhw} at 5.02\,TeV.

Since typical thermalization times used at the LHC ($\sim$0.2~fm/$c$) are comparable to, or even smaller than, the inverse $Y$ binding energies, bound-state formation times are likely to play a role. Within our semiclassical approach, we follow the standard procedure~\cite{Gerland:1998bz} of mimicking the quantum evolution of the forming wave packet as a reduction in the inelastic reaction rate, growing linearly in time from a pointlike $\bbb$ to its asymptotic value as $\tau/\tau_{\rm form}$ for $\tau \leq \tau_{\rm form}$.
The quantum formation times are estimated based on the vacuum binding energies as $\tau_{\rm form} = C/ E_B$, where the coefficient reflects the uncertainty within this approximation; \eg, we find
that a variation of $C$ between 1 and 2 causes a  $\sim$10\% difference in the final results for the $3S$ suppression in peripheral collisions where the formation time is most relevant. 

Finally, after evaluating the number of states that survive traversing the QGP we account for late-time feed-down from excited $Y$ states, following Ref.~\cite{Islam:2020bnp}. 

\begin{figure}[t]
	\centering 
	\includegraphics[width=0.4\textwidth]{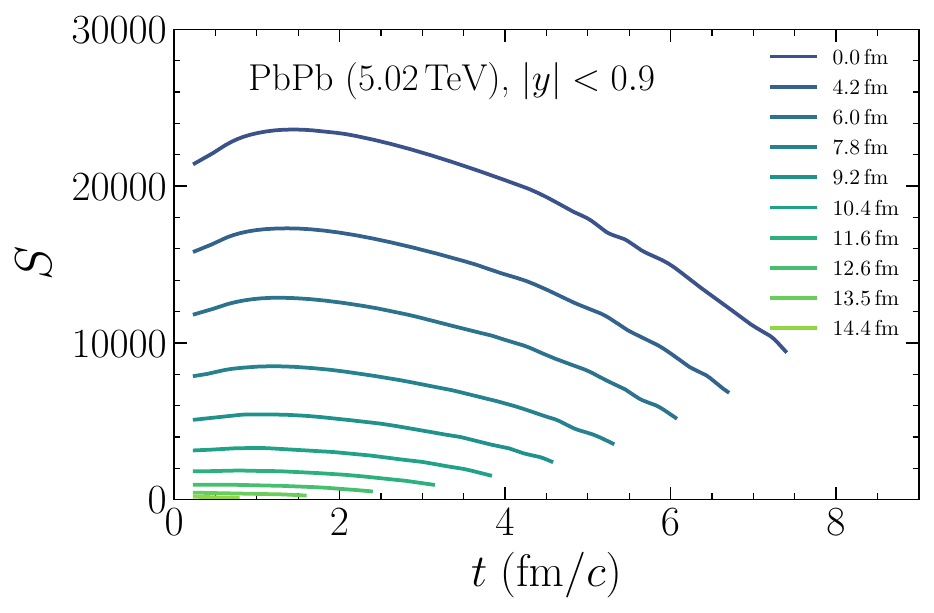}	
	\hfill
	\includegraphics[width=0.4\textwidth]{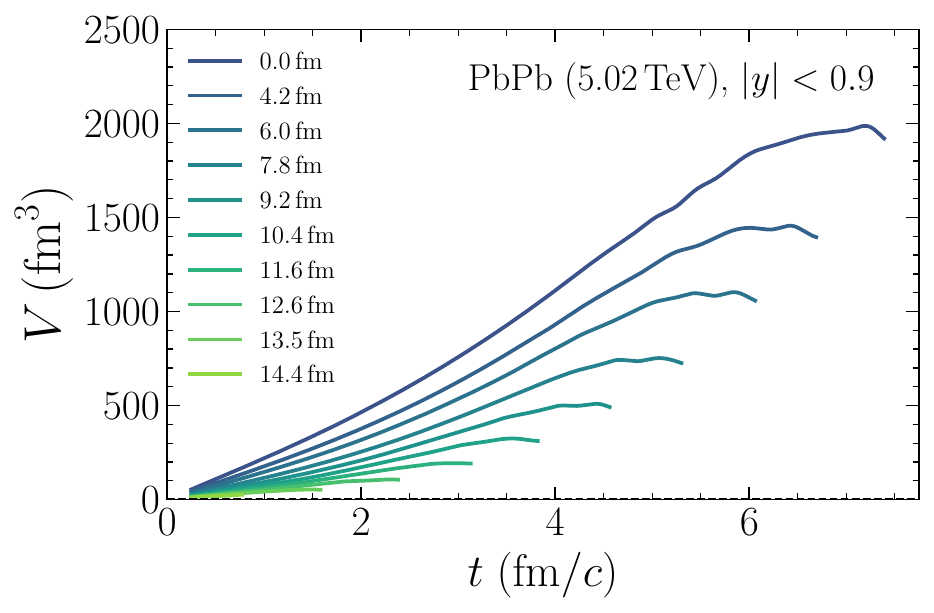}	
	\caption{
	    Time evolution of entropy and volume in hydrodynamics in Pb-Pb collisions at
            $\sqrt{s_{_{\mathrm{NN}}}} = 5.02\,$TeV with $|y| < 0.9$.
    \textbf{Top panel:} Total entropy, $S(t)$, of all fluid elements with $\Tf> 170$\,MeV, as a function of time for different centralities, labeled by the impact parameter $b$.
    \textbf{Bottom panel:} Corresponding volume, $V(t)$, of the medium within the rapidity window.
     } 
        \label{fig:entropy}%
\end{figure}

\section{Regeneration in hydrodynamics}
\label{sec:kinetic}
To evaluate regeneration within a hydrodynamic background with space-time-dependent temperature and flow, we 
employ a thermodynamically-motivated approach, building on our previous applications in an isotropic medium~\cite{Du:2017qkv}.
Alternatively, one could pursue a coupled transport approach of diffusing $b$ and $\bar{b}$ quarks with the kinetics of $Y$ suppression and regeneration. Efforts in this direction have been undertaken~\cite{Yao:2018nmy,Yao:2018sgn,Du:2022uvj}, but a nonperturbative implementation remains to be developed.

The challenge is to evaluate the additional transport parameter, \ie, the equilibrium limit, $N_{Y}^{\rm eq}\left(T(\tau)\right)$, figuring in the gain term of the kinetic rate equation,
\begin{equation}
   \begin{aligned}
      \frac{\dd N_{Y}(\tau)}{\dd\tau} = -\Gamma_{Y}\left(T(\tau)\right) \left[ N_{Y}(\tau) - \Neq \left(T(\tau)\right) \right] \ ,
   \end{aligned}
\label{eq:rate_eq}
\end{equation}
for each $Y$ state in a spatially non-uniform flowing medium.
%
We neglect (``off-diagonal") transitions between bound states as they require color-singlet exchanges which are subleading in deconfined matter; in addition, our reaction rates in the sQGP are large and rapidly equilibrate the excited states while the in-medium feeddown to the ground state is thermally suppressed.
For a uniform fireball at temperature $T$ and volume $V_{\rm FB}$ one has 
\begin{equation}
N_{Y}^{\rm eq}
   =\gamma_b^2 V_{\rm FB}\frac{d_{Y}}{2 \pi^2} T m_{Y}^2  K_2\left(\frac{m_{Y}}{T}\right) \ ,
   \label{eq:density}
\end{equation}
with spin-degeneracy $d_Y$, mass $m_{Y}$ and modified Bessel function of the second kind, $K_2$.
The fugacity $\gamma_b=\gamma_{\bar b}$ is fixed by the conservation of $\bbb$ pairs in the fireball, 
\begin{equation}
N_{b \bar{b}}=\frac{1}{2} \gamma_b n_{\mathrm{op}} V_{\mathrm{FB}} \frac{I_1\left(\gamma_b n_{\mathrm{op}} V_{\text {corr }}\right)}{I_0\left(\gamma_b n_{\mathrm{op}} V_{\text {corr }}\right)}+\gamma_b^2 n_{\mathrm{hid}} V_{\mathrm{FB}}\,,
\label{eq:fugacity}
\end{equation}
with open- and hidden-bottom densities $n_{\rm op}$ and $n_{\rm hid}$, respectively, and modified Bessel functions of the first kind, $I_{0,1}$. In the QGP, we include $b$-quarks and the $S$-wave meson resonances in $n_{\rm op}$ as in Ref.~\cite{Du:2017qkv}, but further add $S$-wave baryon states~\cite{Kaczmarek:2025dqt} (the latter reduce $\Neq$ equilibrium limit by about $\sim$10\%). 

In the canonical ensemble, the correlation volume, $V_{\rm corr}$, in $I_0/I_1$ ensures exact $\bbb$ conservation starting from an initial production volume of the pair. Its time evolution is taken as~\cite{Grandchamp:2003uw} 
\begin{equation}
V_{\rm corr}
= \tfrac{4}{3}\pi\bigl(r_0 + \langle v_b\rangle\,\tau\bigr)^3\,,
\end{equation}
where $r_0=1$\,fm characterizes an initial strong-interaction range, and $\langle v_b\rangle=0.65\,c$ the average $b$-quark recoil velocity estimated from $B$-meson $\pT$ spectra. The sensitivity of the regeneration yield to the parameterization of this volume turns out to be rather weak~\cite{Du:2017qkv}.

Incomplete $b$-quark thermalization is accounted for in a relaxation time approximation~\cite{Grandchamp:2002wp}, 
via a correction factor,
\begin{equation}
\mathcal{R}(\tau)
= 1 - \exp\left[-\!\int_{\tau_0}^{\tau}\frac{\dd\tau'}{\tau_b}\right] \ ,
\end{equation}
which has been shown to produce fairly reliable results~\cite{Song:2012at}. Non-thermal $b$-quark spectra are less favorable for bound-state formation and thus reduce regeneration. The pertinent relaxation time, $\tau_b\simeq7.5$\,fm/$c$, is calculated 
from the same interactions as the $Y$ reaction rates~\cite{Tang:2023tkm}.

To define the ``active" fireball volume, $V_{\rm FB}=V_{\rm act}$, of the hydrodynamic medium in which regeneration can take place, we approximate the local environment by the trajectory-averaged temperature, $T_{\rm avg}(\tau)$, sampled along the $Y$ paths as shown in Fig.~\ref{fig:traj}. The active volume is then  evaluated from the total entropy, $S(\tau)$, and local entropy density, $s(T)$, in the hydro cell, 
\begin{equation}
V_{\rm act}(\tau)
= \frac{S(\tau)}{s\bigl[T_{\rm avg}(\tau)\bigr]}\,,
\end{equation}
with
\begin{equation}
S(\tau)
= \int_{T(\tau,\mathbf{x})>T_{\rm f}}\!\dd^3x\; s\bigl[T(\tau,\mathbf{x})\bigr]\, ;
\end{equation}
the integration is over the hydrodynamic eigenvolume for temperatures larger than the freeze-out temperature $\Tf$.
In Fig.~\ref{fig:entropy} we display $S(\tau)$ and the resulting $V_{\rm act}(\tau)$ in Pb-Pb (5\,TeV) collisions at mid-rapidity for various centralities with $\Tf$=170\,MeV (as in our treatment of suppression). After an initial increase in the build-up phase, the system cools and $S(\tau)$ decreases more rapidly while $V_{\rm act}(\tau)$ grows, reflecting the fireball's expansion. 

\begin{figure*}[t]
	\centering 
	\includegraphics[width=0.32\textwidth]{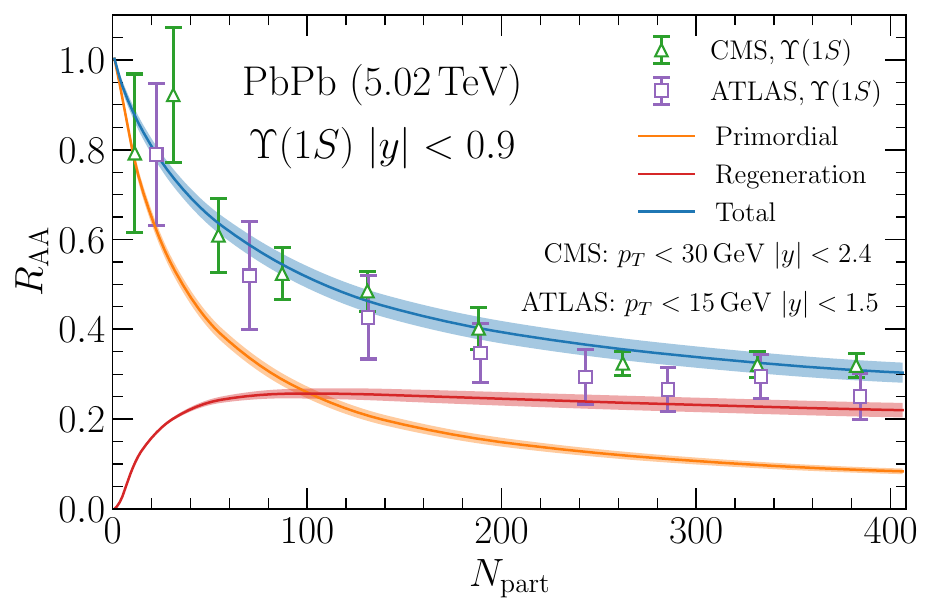}	
	\includegraphics[width=0.32\textwidth]{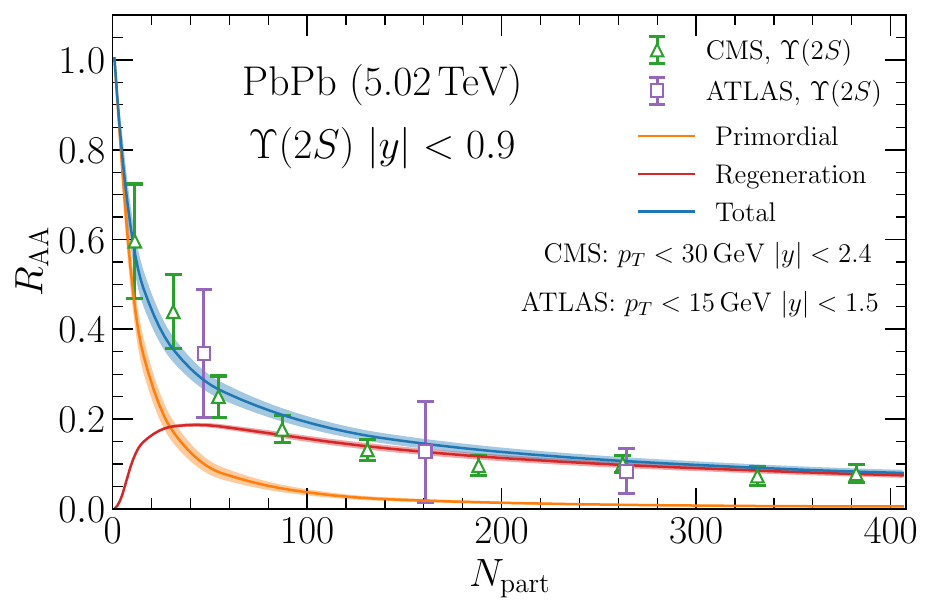}	
	\includegraphics[width=0.32\textwidth]{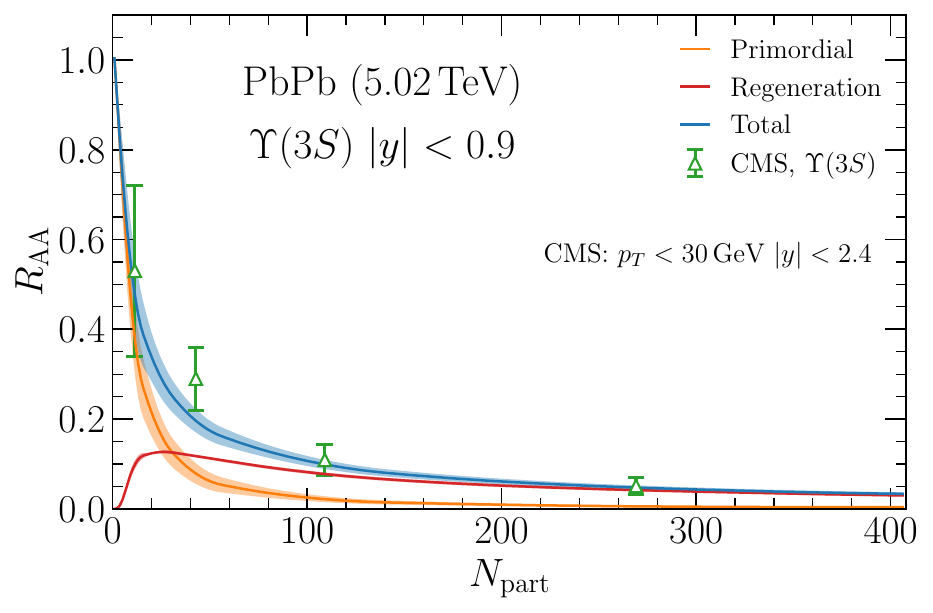}	
	\includegraphics[width=0.32\textwidth]{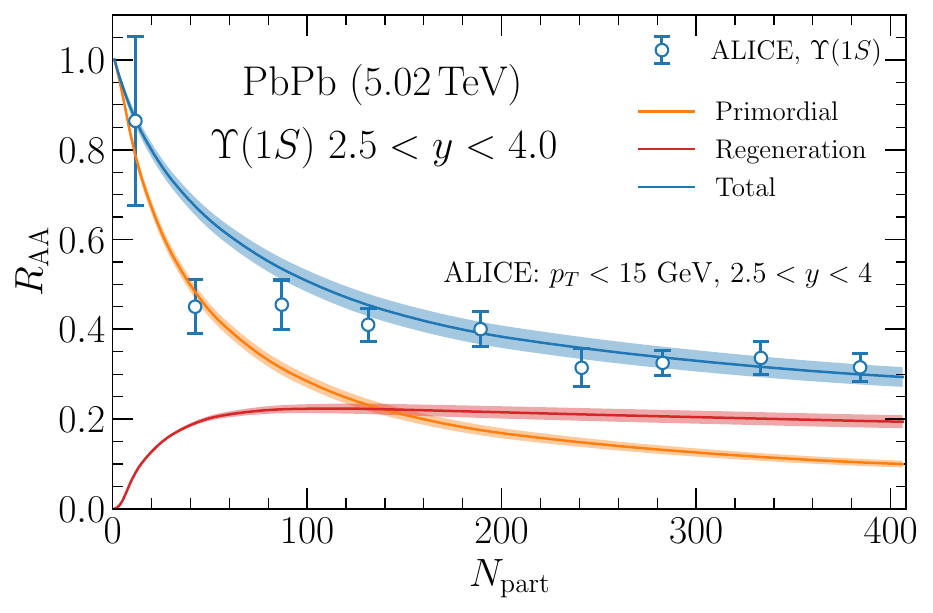}	
	\includegraphics[width=0.32\textwidth]{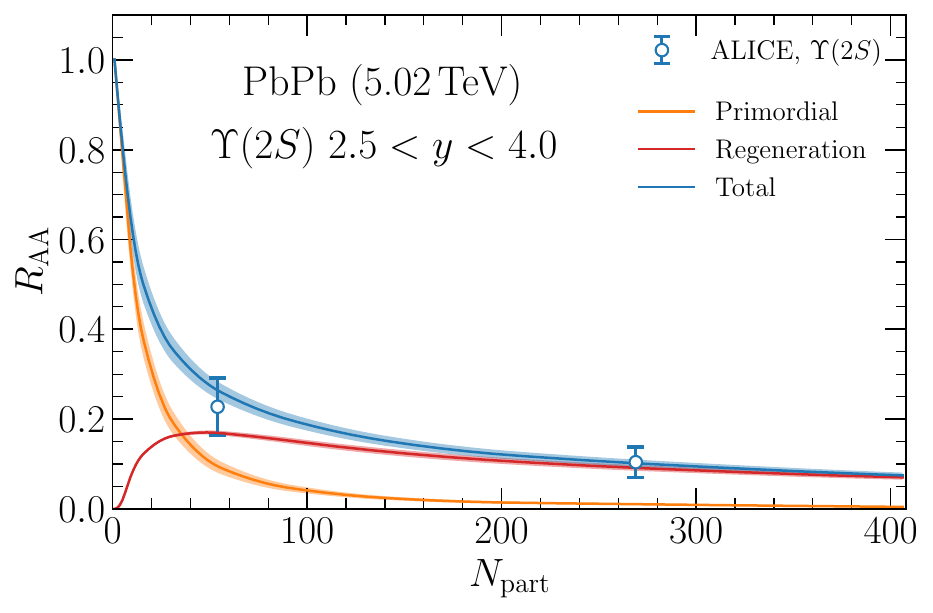}	
	\includegraphics[width=0.32\textwidth]{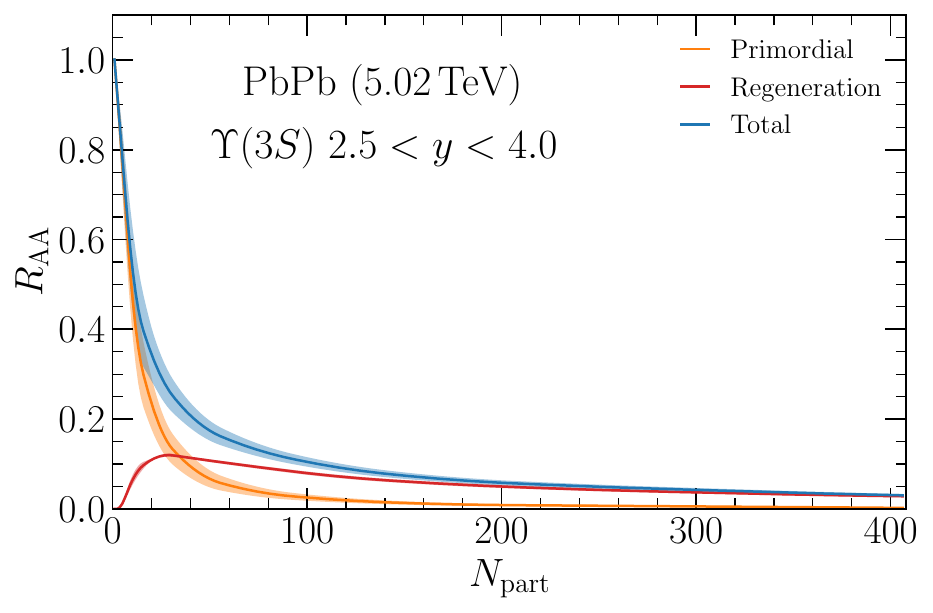}	
	\caption{
	   $R_{AA}$ of $\Upsilon(1S)$ (left), $\Upsilon(2S)$ (middle), and $\Upsilon(3S)$ (right) as a function of the number of participants at mid-rapidity (upper) and forward rapidity (lower) compared to the CMS data~\cite{CMS:2018zza,CMS:2023lfu}, ATLAS data~\cite{ATLAS:2022exb}, and ALICE data~\cite{ALICE:2018wzm,ALICE:2020wwx}.
    The bands represent our results with a shadowing of 20\%--30\% (mid-rapidity) and of 25\%--35\% (forward-rapidity) together with the uncertainties from the bottomonium formation time, $\tau_{\rm form}$.
  }
	\label{fig:npart}%
\end{figure*}

The total $\bbb$-pair number is set by the production cross section and the inelastic cross section in $pp$ collisions, multiplied by the number of binary collisions $N_{\rm coll}$ for a given centrality and rapidity bin, $\Delta y=1.8$, 
\begin{equation}
N_{b\bar b}
= \frac{\dd\sigma_{b\bar b}}{\dd y}\;\frac{N_{\rm coll}}{\sigma_{pp}^{\rm inel}}\;\Delta y \ , 
\end{equation}
and corrected for shadowing as described above.

Much like for the bottomonia, $b$ quarks may exit the active fireball volume before the temperature has dropped below $T_f$.
We estimate the fraction of $b\bar{b}$ pairs that remain within the fireball at each time step,
\begin{equation}
f_{b\bar{b}}(\tau) = \frac{N_{b\bar{b}}^{\text{in}}(\tau)}{N_{b\bar{b}}^{\text{total}}} \, , 
\end{equation}
by sampling their initial positions according to a Glauber-model distribution
and their $\pT$ from FONLL distributions, assuming back-to-back kinematics.
Initially, all $b\bar{b}$ pairs are inside the fireball (except for very peripheral collisions where they have a finite probability to be produced outside the region with $T\ge T_f$). 
By applying this fraction to the total number of $b\bar{b}$ pairs, we obtain the effective number of $b\bar{b}$ pairs present in the fireball at each time step,
\begin{equation}
N_{b\bar{b}}^{\text{eff}}(\tau) = f_{b\bar{b}}(\tau) \cdot N_{b\bar{b}}(\tau) \, .
\end{equation}

Regeneration processes become operative once the medium has cooled to the relevant melting temperature, $\Tm^Y$, for each state.
In our earlier works, we assumed this temperature to be defined by the vanishing of the pertinent binding energies, $E_B^Y$~\cite{Du:2017qkv}.
However, in recent work~\cite{Tang:2025ypa} it was discovered (by using a complex-pole analysis of the bottomonium $T$-matrices) that 
resonance-like correlations can persist even beyond the temperature where the nominal binding energy vanishes.
Therefore, we have investigated two different scenarios for the onset of regeneration (defined by the lower limit of the integral, $\tm$, in the gain term of Eq.~\eqref{eq:rate_eq}), namely $\tm=\tau(E_B=0)$ at which $E_B^Y$ vanishes and $\tm=\tau(T=\Tm)$ when pole of the $T$-matrix disappears. It turns out that the resulting differences in the final regeneration yields are very small.

\section{Predictions for LHC $Y$ data}
\label{sec:data}
In our comparisons to experimental data we include error bands where the theoretical uncertainties due to nuclear shadowing and $Y$ formation times, $\tau_{\rm form}^Y$ (as discussed above) are added in quadrature.
We focus on Pb-Pb collisions at $\sqrt{s_{_{\rm NN}}}=5.02\,$TeV, specifically the nuclear modification factor
\begin{equation}
\raa^Y(N_{\rm part}) = \frac{N_Y^{\rm AA}(N_{\rm part};\pT)}{N_Y^{pp}(\pT) \,N_{\rm coll}(N_{\rm part})}\,,
\label{eq:raa}
\end{equation}
where $N_Y^{\rm AA}(\npart)$ is the inclusive $Y$ yield measured in nucleus-nucleus collisions for a given number, $\npart$ of participating nucleons, $N_Y^{pp}(\pT)$ is the corresponding yield in $pp$ collisions at the same collision energy,
and $N_{\rm coll}(\npart)$ is the number of primordial binary nucleon--nucleon collisions for a given centrality selection.
We discuss the centrality dependence in Sec.~\ref{ssec:npart}, followed by the $\pT$ spectra in Sec.~\ref{ssec:pT}.

\subsection{Centrality dependence}
\label{ssec:npart}
Figure~\ref{fig:npart} displays our results for $\raa(\npart)$ for $\Upsilon(1S)$, $\Upsilon(2S)$, and $\Upsilon(3S)$ at both mid- and forward rapidity, compared to 
CMS~\cite{CMS:2018zza,CMS:2023lfu}, ATLAS~\cite{ATLAS:2022exb}, and ALICE~\cite{ALICE:2018wzm,ALICE:2020wwx} data.
\begin{figure*}[t]
	\centering 
	\includegraphics[width=0.32\textwidth]{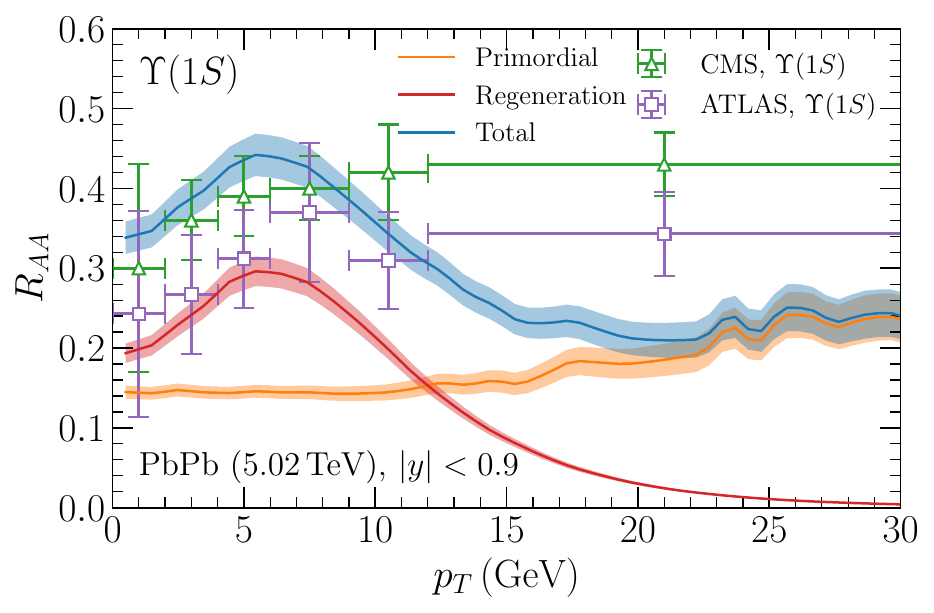}	
	\includegraphics[width=0.32\textwidth]{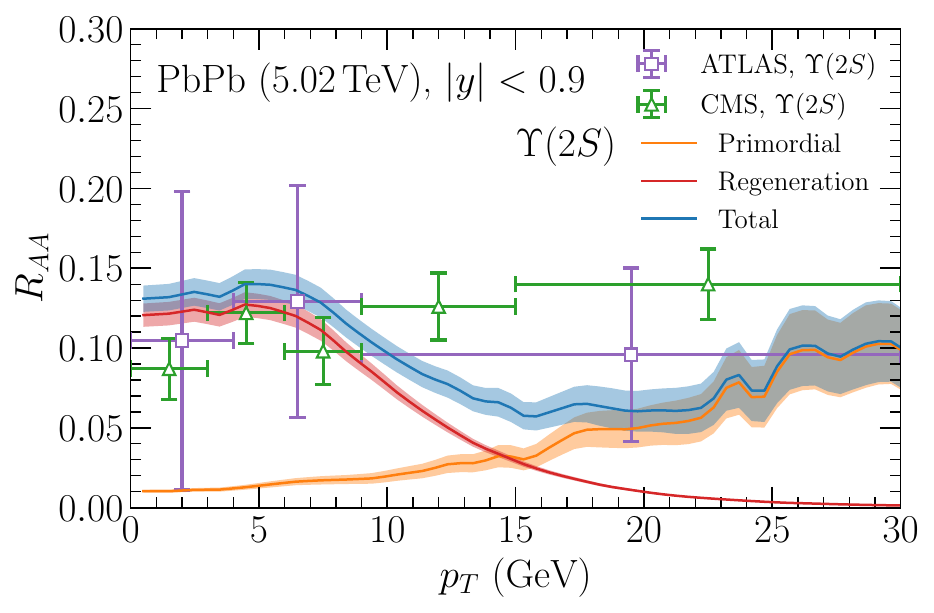}	
	\includegraphics[width=0.32\textwidth]{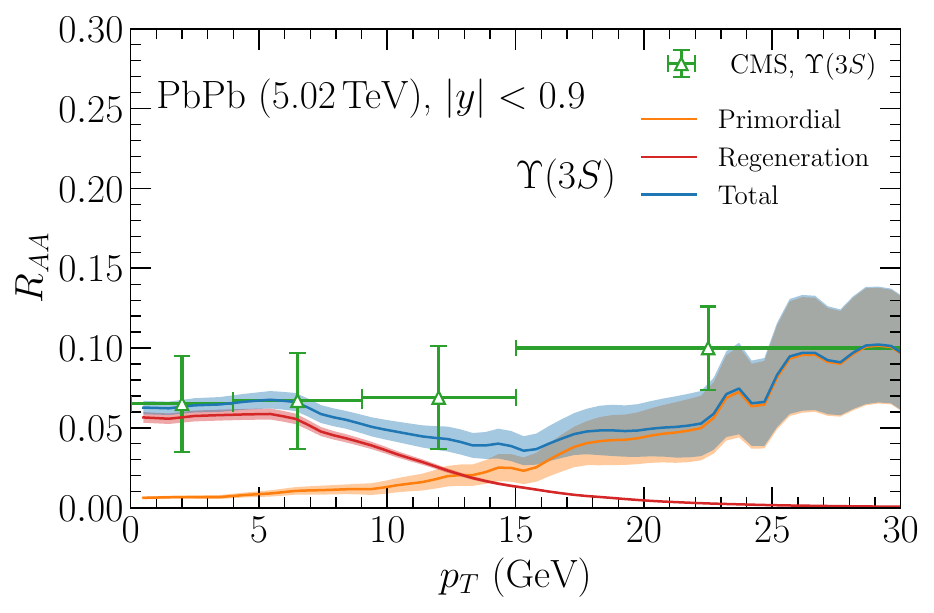}	
	\caption{
	   Minimum-bias $\pT$-spectra of the total (blue), primordial (orange) and regenerated (red) 
    $\Upsilon(1S)$ (left), $\Upsilon(2S)$ (middle), and $\Upsilon(3S)$ (right) at mid-rapidity in Pb-Pb (5.02\,TeV) collisions.
    The bands represent a 20-30\% shadowing range, added in quadrature with the uncertainties from the bottomonium formation times.
 }
	\label{fig:pt}%
\end{figure*}
At mid-rapidity, displayed in the upper panel, our predictions for $\Upsilon(1S)$, $\Upsilon(2S)$ and $\Upsilon(3S)$ (left, middle and right panel, respectively) agree fairly well with CMS and ATLAS data.  
The quality of the data description is comparable to our previous calculations based on a perturbative coupling to a quasiparticle QGP~\cite{Du:2017qkv}. However, the present description is conceptually superior since the calculated rates are rooted in constraints from lattice-QCD and selfconsistently coupled to a strongly interacting medium with broad parton spectral functions representing large collision rates in the QGP liquid. The composition of the yields is also different: the larger $Y$ reaction rates imply significantly more suppression but also more regeneration. The latter becomes the leading contribution for $\Upsilon(1S)$ in central collisions, while it is the dominant source of $\Upsilon(2S)$ and $\Upsilon(3S)$ starting from rather peripheral collisions, for $\npart\gtrsim50$. 

This picture essentially persists at forward rapidity, where the partitioning between suppressed primordial and regenerated production is very similar to mid-rapidity, yielding fair agreement with ALICE data except for a tension in peripheral collisions where $\Upsilon(1S)$ production tends to be over-predicted.

\subsection{Transverse-momentum spectra}
\label{ssec:pT}
While the predictions for the inclusive yields with the nonperturbative rates seem to work out reasonably well, they do not render a decisive signature compared to other scenarios. But a larger regeneration yield may leave more distinct features in the $\pT$ spectra. A reliable calculation of the latter requires to account for the non-equilibrium  spectra of the $b$ quarks as they diffuse through the fireball (as mentioned above), which we have thus far not explicitly accounted for. Here, we approximate the $Y$ $\pT$ spectra by an instantaneous coalescence model (ICM)~\cite{Du:2017qkv}
\begin{eqnarray}
\frac{\dd^{3} N_{Y}^{\mathrm{reg}}\left(\mathbf{p}\right)}{\dd^{3} \mathbf{p}} =
C_{\mathrm{reg}}d_{Y}\int \dd^{3} \mathbf{p}_{b} \dd^{3} \mathbf{p}_{\bar{b}} \frac{\dd^{3} N_{b}}{\dd^{3} \mathbf{p}_{b}} \frac{\dd^{3} N_{\bar{b}}} {\dd^{3} \mathbf{p}_{\bar{b}}}
\nonumber \\
\hfill{ } \qquad  \times \,\delta^{(3)}\left(\mathbf{p}-\mathbf{p}_{b}-\mathbf{p}_{\bar{b}}\right)w\left( \mathbf{k} \right)\ .
\end{eqnarray}
We employ transported $b$-quark spectra, $\dd^3N_b/\dd^{3} \mathbf{p}_{b}$, from HQ diffusion simulations~\cite{He:2014cla} at an average regeneration temperature for each $Y$ state, \ie, T=280, 180, 170\,MeV for $\Upsilon(1S)$, $\Upsilon(2S)$ and $\Upsilon(3S)$, respectively. In the Wigner function, $w(k)$, we insert the same radii as used in the interference factor of the reaction rates from Ref.~\cite{Wu:2025hlf}. The overall normalization, $C_{\rm reg}$, is fixed to the values obtained from the rate equation at the end time, $\tau(\Tf)$.
The ICM results are combined with the $\pT$-dependent suppressed primordial component into the $\raa(\pT)$ for minimum bias Pb-Pb collisions for inclusive $1S$, $2S$, and $3S$ bottomonia at mid-rapidity and compared
to CMS~\cite{CMS:2018zza,CMS:2023lfu}, 
and ATLAS~\cite{ATLAS:2022exb} data in Fig.~\ref{fig:pt}.
For all 3 states, a maximum structure at relatively low $\pT\lesssim10$\,GeV develops which is most pronounced for the $\Upsilon(1S)$, less so for the $\Upsilon(2S)$ and essentially only a shoulder for the $\Upsilon(3S)$. 
The data are not inconsistent with these structures but do not show compelling evidence in favor of them either. 
At high $\pT$ our calculations tend to underestimate the data for all 3 states, even though the primordial component increases at higher $\pT$ due to smaller reaction rates and time-dilated formation time effects. There could be several reasons for that, 
\eg, longer thermalization times for the onset of the hydro evolution especially in peripheral collisions (currently assumed to be constant at $\tau_0$=0.25\,fm)
or the onset of different production mechanisms~\cite{Aronson:2017ymv} (\eg, gluon splitting~\cite{Bedjidian:2004gd}) in connection with an explicit quantum evolution~\cite{Blaizot:1987ha}, which might reduce the suppression of the primordial component. Also, space-momentum correlations between the $b$ and $\bar b$ quarks, not included here, are known to extend the relevance of regeneration processes~\cite{He:2021zej}. 


\section{Summary}
\label{sec:sum}
We have presented a new calculation of quarkonium transport in heavy-ion collisions within a fully integrated nonperturbative framework. The two main building blocks are recently obtained microscopic reaction rates based on lattice-QCD constraints and a viscous hydrodynamic medium evolution tuned to light-hadron data. This merges the nonperturbative micro-physics of bound-state structure and medium coupling with the macro-physics of a hydrodynamic evolution through the same equation of state in a selfconsistent way, while also establishing consistency with the nonperturbative interactions required to describe the individual diffusion of heavy quarks in the sQGP. 
While both building blocks did not involve tunable parameters, their current implementation still involves a number of modeling components, \eg, the quarkonium equilibrium limit in evaluating regeneration processes in a spatially non-uniform medium, and the concepts of bound-state formation times and melting temperatures in the early evolution to mimic quantum effects in our semiclassical treatment. We have elaborated them via parameter variations and included the most relevant ones in our final results.
We have estimated uncertainties from variations in our input parameters, most notably nuclear shadowing in the primordial production of $Y$'s and $b\bar b$ pairs affecting initial conditions and the magnitude of regeneration. Both turned out to be rather modest as part of our theoretical error estimates.

In our applications to experiment we have focused on bottomonium production in Pb-Pb collisions at the LHC. The agreement with the centrality dependence of various $Y$ states turned out to be comparable to previous fireball calculations with perturbative coupling to a quasiparticle medium. However,the much larger nonperturbative reaction rates predict a marked change in the composition of the $Y$ yields, with stronger suppression and a marked increase of regeneration yields. Consequently, the latter become the leading source of $\Upsilon(1S)$ states in semi-/central collisions, while it dominates $\Upsilon(2S)$ and $\Upsilon(3S)$ production starting from peripheral collisions on.  
This leaves significant footprints in the transverse-momentum spectra, with a maximum in the nuclear modification factor at low $\pT$ while toward high $\pT$, our results are somewhat below the data. An even stronger footprint might come from the $v_2$ spectra. We still expect the $v_2$ of the $\Upsilon(1S)$ to be fairly small since its kinetics mostly occurs early in the fireball evolution. However, for all excited states the $v_2$ should be quite substantial, since their production is dominated by regeneration relatively late in the evolution where the $b$ quarks are likely to have developed appreciable collectivity.

We envisage a number of future directions and improvements.
Clearly, the discrepancy at high momentum needs to be addressed.
It will also be critical to test how the nonperturbative micro-/macro-approach fares for charmonia and $B_c$ mesons.
A quantum transport treatment for the earlier phases is certainly in order; our results suggest that it is vital to accomplish that with a realistic regeneration mechanism encompassing all $b\bar{b}$ pairs in the system, which thus far is not available. Furthermore, a more microscopic treatment of regeneration is in order, to improve the control over the $\pT$ dependence in the recombination of the diffusing heavy quarks, in particular also for the elliptic flow. Work in some of these directions is in progress.


\section*{Acknowledgments}
We thank Jacob Boyd, Sabin Thapa and Ramona Vogt for valuable discussions. This work is supported by the U.S. National Science Foundation under grant no.\,PHY-2209335 and the Department of Energy via the Topical Collaboration in Nuclear Theory on \textit{Heavy-Flavor Theory (HEFTY) for QCD Matter} under award no.\,DE-SC0023547.


\bibliographystyle{apsrev4-1}

\bibliography{refcnew}






\end{document}